\documentclass[aps,prl,twocolumn,showpacs,superscriptaddress,groupedaddress]{revtex4}  

\usepackage{graphicx}
\usepackage{dcolumn}
\usepackage{bm}
\usepackage{amssymb}
\usepackage{epstopdf}
\usepackage{amsmath}
\usepackage{amsfonts}
\usepackage{color}
\usepackage{mathrsfs}
\usepackage{comment}

\newcommand{\be}{\begin{equation}}
\newcommand{\ee}{\end{equation}}
\newcommand{\bea}{\begin{eqnarray}}
\newcommand{\eea}{\end{eqnarray}}
\newcommand{\barr}{\begin{array}}
\newcommand{\earr}{\end{array}}

\newcommand{\thetabubble}{\theta_{\rm bubble}}

\def\n{{\bf \hat n}}

\def\beq{\begin{equation}}
\def\eeq{\end{equation}}
\def\be{\begin{equation}}
\def\ee{\end{equation}}
\def\bea{\begin{eqnarray}}
\def\eea{\end{eqnarray}}

\def\n{{\bf \widehat n}}

\def\ba{\begin{align}}
\def\ea{\end{align}}

\def\L{{\mathcal L}}

\begin{document}

\title{Collisions with other Universes: the Optimal Analysis of the WMAP data}

\author{S.J.~Osborne}
\affiliation{Department of Physics, Stanford University, Stanford, CA 94306}
\affiliation{Kavli Institute for Particle Astrophysics and Cosmology, Stanford University and SLAC, Menlo Park, CA 94025}
\author{L.~Senatore}
\affiliation{Department of Physics, Stanford University, Stanford, CA 94306}
\affiliation{Kavli Institute for Particle Astrophysics and Cosmology, Stanford University and SLAC, Menlo Park, CA 94025}
\affiliation{CERN, Theory Division, 1211 Geneva 23, Switzerland}
\affiliation{Stanford Institute for Theoretical Physics, Stanford University, Stanford, CA 94306}
\author{K.M.~Smith}
\affiliation{Department of Astrophysical Sciences, Princeton University, Princeton, NJ 08544-1001, USA}
\affiliation{Perimeter Institute for Theoretical Physics, Waterloo, ON N2L 2Y5, Canada}

\date{\today}

\begin{abstract}
An appealing theory is that our current patch of universe was born
as a nucleation bubble from a phase of false vacuum eternal inflation.
We search for evidence for this theory by looking for the signal imprinted on
the CMB that is generated when
another bubble ``universe'' collides with our own.
We create an efficient and optimal estimator for the signal in the
WMAP 7-year data.
We find no detectable signal, and constrain the amplitude, $a$, of the
initial curvature perturbation that would be generated by a collision:
$-4.66 \times 10^{-8} < a \left( \sin{\thetabubble}\right)^{4/3} < 4.73 \times 10^{-8} \; [{\rm Mpc}^{-1}]$
at 95\% confidence
where $\thetabubble$ is the angular radius of the bubble signal.
\end{abstract}

\maketitle


{\em Introduction:}  Few things are more exciting than discovering what happened at the beginning of the universe, or 
understanding the structure of spacetime outside our observable universe.
Quite remarkably, there exist cosmological signatures that would allow us to address exactly these questions. In this paper we concentrate on one of these, which is the signature imprinted on the Cosmic Microwave Background radiation (CMB) by primordial bubble collisions.  
An appealing theory for the origin of the universe is that it was created by quantum mechanical tunneling from a much larger eternally inflating spacetime. This larger spacetime expands exponentially, driven by the energy of the false vacuum. Occasionally one region of space will tunnel to a lower energy vacuum. Even though the resulting bubble-like region expands at the speed of light, and bubbles are continuously produced in many places, the false vacuum region expands  quickly enough that the bubbles never percolate. This is the so-called False Vacuum Eternal Inflation~\cite{Guth:1980zm}. In this scenario, we live in the interior of one of the bubbles, where our standard slow roll inflationary phase
takes place.
Although bubbles do not percolate and fill the whole of space, there is a chance that in the past another
bubble collided with our own~\cite{Freivogel:2009it}, leaving a specific disk-shaped signal in the CMB~\cite{Chang:2008gj}. This is a localized and well defined
signature. Discovery of such a collision would have tremendous implications for the whole field of high-energy
physics. We would learn of a new cosmological epoch, eternal inflation, that happened before the epoch of standard
inflation. Furthermore, we would learn  that the field theory that describes the universe has at least two vacua,
one false and one true. This would already be an important discovery. In addition, the detection of a bubble
collision would hint that there are many other universes like ours, a whole landscape of vacua. The presence
of a landscape of vacua would provide indirect evidence of string theory, which predicts such
a landscape, and also of Weinberg's anthropic explanation of the cosmological constant~\cite{Weinberg:1987dv},
which relies on a landscape of vacua each with a different vacuum energy.  
This motivates us to search for bubble collisions in the WMAP data.
Implementing an exact, fully optimal likelihood analysis is still not a completely solved problem,
although considerable progress has been made recently in~\cite{Feeney:2010dd,McEwen:2012sv,Feeney:2012hj}.
In the companion paper~\cite{technical}, we introduce new algorithmic tricks which
solve this problem, making the optimal analysis computationally affordable and
furthermore simplifying the methodology.
Our tools are particularly powerful for bubbles with a sharp edge feature (the ``step'' profile defined below),
where the typical bubble is large but the CMB maps must be kept at high resolution, since most of the signal-to-noise
comes from small angular scales.
For bubbles without a sharp edge feature (the ``ramp'' profile below), the statistical weight comes
entirely from angular scales of order 1 degree or larger, where WMAP is sample variance limited and
additional data (e.g.~from Planck) will not improve the measurement.
The statistically optimal constraints reported in this paper therefore represent the ultimate
constraints which can be obtained for this profile using CMB temperature.
Finally, we extend existing analysis techniques by complementing the Bayesian analysis by
a Monte Carlo based approach which does not depend on an external prior
for the bubble amplitude.
In this paper we focus on the results and describe the technical details in~\cite{technical},
where we highlight how the analysis techniques that we develop in this context can be applied
to all localised features in the CMB.
We use the WMAP~7-year cosmological parameters throughout~\cite{Komatsu:2010fb}.

\smallskip
{\it Bubble Signal:} The theory of bubble collisions has recently been
reviewed in~\cite{Kleban:2011pg}.
While the amplitude of the signal depends on the details of
the dynamics describing the collision, the shape of the signal does not,
and it appears as a disk on the sky. 
We model the bubble collision as a perturbation to the initial adiabatic
curvature $\zeta({\bf x})$, and consider two possible forms:
either a ``ramp'' profile
$\zeta_r(z) =a^{\rm ramp} (z-r)$ for $z \ge r$
and zero for $z \le r$ (where $r$ is the comoving distance to the bubble wall),
or a ``step'' profile
$\zeta_r(z) = a^{\rm step}$ for $z \ge r$ and zero for $z \le r$.
We propagate this curvature perturbation to a CMB temperature perturbation
using the full numerically computed CMB transfer function.
As described in detail in~\cite{technical}, the transfer function
corrects the ramp profile by $\approx 10$\% and the step profile at order unity,
so including it is necessary for a precise analysis.

The theoretical distribution of bubble sizes is determined by the symmetry and geometry
of the bubble collision, and we incorporate this information into our analysis.
For small curvature $\Omega_k$, the comoving distance $r$
to the bubble wall is uniformly distributed~\cite{Kleban:2011pg}.
Equivalently, the size distribution is given by
$dP(\thetabubble) \propto d\cos{\thetabubble}=\sin{\thetabubble} \, d\thetabubble$.
The ``typical'' bubble is large; its angular size is of order one radian.
For the ramp profile, most of the signal for detecting such a bubble comes from
angular scales comparable to the bubble radius (roughly $\ell\lesssim 20$), where WMAP is cosmic
variance limited.
For the step profile, the signal comes mainly from high-$\ell$ modes associated with
the sharp edge feature, and Planck can potentially improve the optimal WMAP constraints
presented here.


\smallskip
{\em Method:} We search for the bubble signature in the WMAP
7-year V-band and W-band data.
To minimize foreground contamination, we use foreground reduced
maps, and apply the WMAP KQ75 extended temperature analysis mask
to exclude the Galaxy and bright point sources.
We use the algorithmic machinery from~\cite{technical} to perform
all-sky exact evaluation of the Bayesian likelihood and optimal
frequentist statistic; we summarize the key steps as follows.

Since the CMB and WMAP noise are Gaussian to a very good approximation, the likelihood
for obtaining data realization $d$ has
the form $\L(d) \propto \exp(-\chi^2(d)/2)$,
where we have defined $\chi^2(d) = d^T C^{-1} d$.
Here, $C$ is the data covariance matrix with dimension $(nN_{\rm pix})^2$, where $n$ and $N_{\rm pix}$
are the number of differencing assemblies (DA's) and pixels.
We use the exact CMB + noise covariance matrix throughout, which leads to optimal statistics by optimally
weighting the maps in the presence of masking, noise inhomogeneity, and per-DA beams.

We consider a single-bubble model, which has four parameters:
the amplitudes $a^{\rm ramp}, a^{\rm step}$ of the ramp and step perturbations,
the distance $r$ to the collision wall, and the direction $\n$ of the bubble center.
All information is contained in the change of $\chi^2$ when the bubble is subtracted 
from the data, defined by
\be
\Delta\chi^2(d,a,r,\n) = \chi^2\!\!\left( d - a^{\rm ramp} \beta^{\rm ramp}_{r,\n} - a^{\rm step} \beta^{\rm step}_{r,\n} \right) - \chi^2(d)
\ee
where $a=(a^{\rm ramp},a^{\rm step})$, and $\beta_{r,\n}$ denotes the bubble profile, including CMB transfer functions and beam convolution.
Calculating $\Delta\chi^2$ would be computationally prohibitive
were it not for several computational efficiencies that we make use of.
The first is the preconditioned conjugate gradient descent
algorithm of~\cite{Smith:2007rg} that allows us to calculate the $C^{-1}$ operation efficiently.
The second is a method that allows us to efficiently calculate $\beta^T_{r,\n} C^{-1} \beta_{r,\n}$
(which naively would require performing the $C^{-1}$ operation for many values of $r$ and for
every pixel $\n$ in the WMAP map).


\smallskip
{\em Bayesian analysis:} 
Our starting point for Bayesian analysis of the bubble collision signal is the posterior
likelihood for model parameters $(a^{\rm ramp}, a^{\rm step}, r)$, marginalized over the
bubble location $\n$:
\be
\label{eqn:bayes_likelihood}
\mathcal{L}(a^{\rm ramp}, a^{\rm step}, r | d)  \propto
\frac{p(r)}{N_{\rm pix}} \sum_{\n} \exp \left(-\frac{1}{2} \Delta\chi^2(d,a, r, \n) \right)
\ee
where $p(r)$ denotes the uniform prior on the distance parameter $r$.

\begin{figure}[t]
  \centering
  \includegraphics[width=71mm, clip=true,trim=0.1cm 0 0 0]{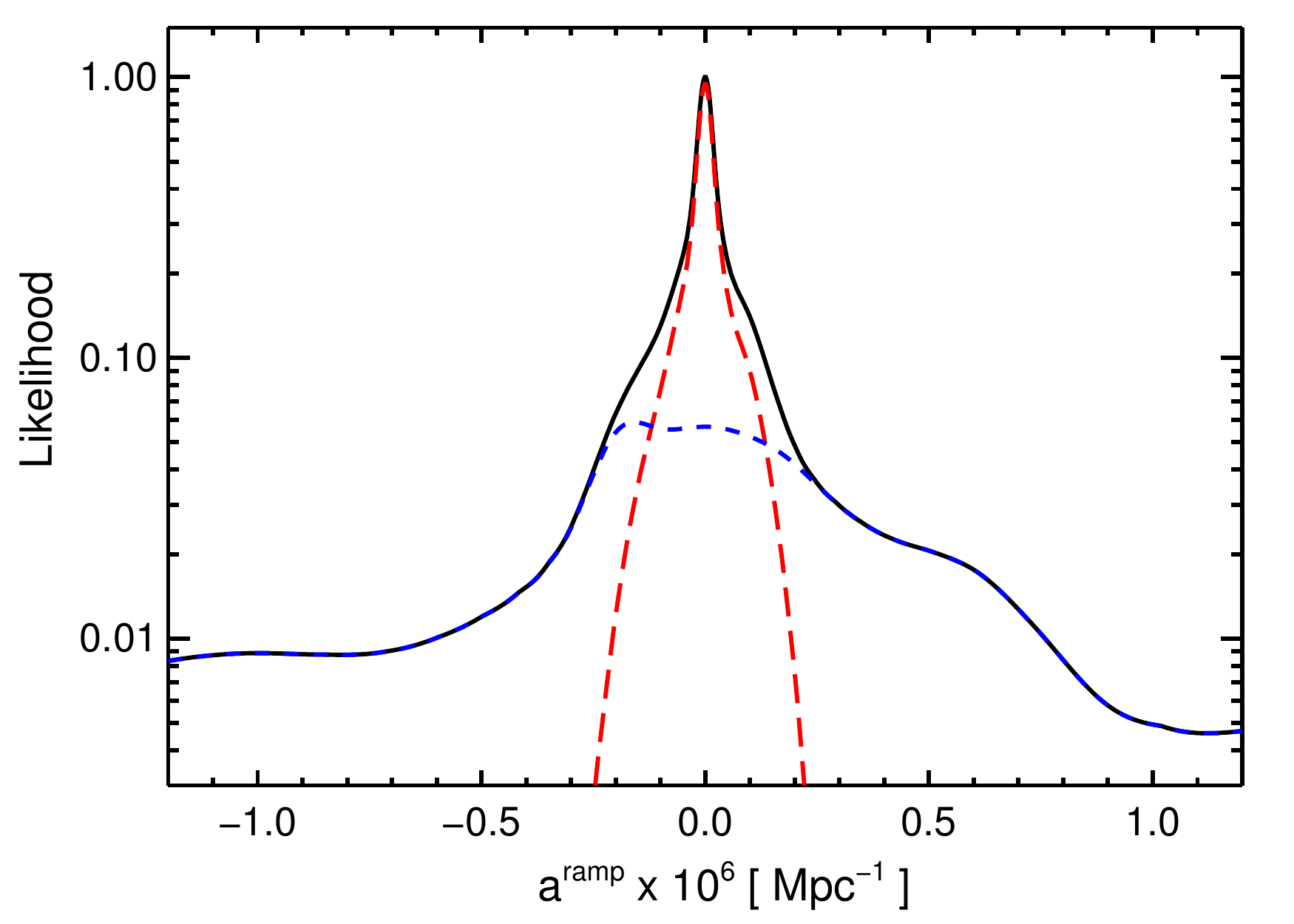}
  \includegraphics[width=71mm, clip=true,trim=0.1cm 0 0 0]{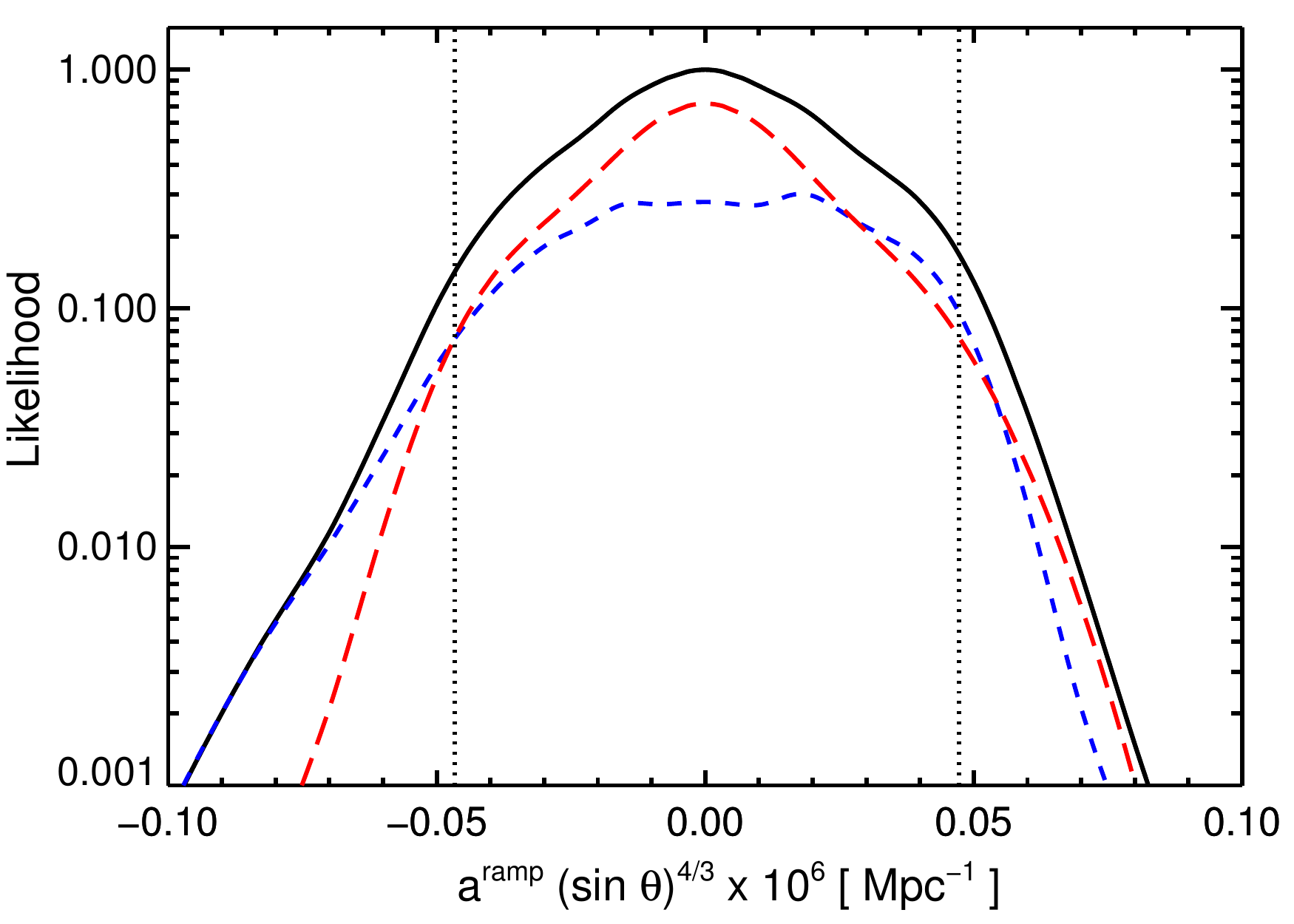}
  \caption{Bayesian analysis of the bubble parameter space, assuming the ``ramp model''
for the bubble profile. \\
{\em Top panel:} Posterior likelihood ${\mathcal L}(a^{\rm ramp}|d)$ for the amplitude
parameter $a^{\rm ramp}$, defined to be the slope of the initial curvature perturbation
in Mpc$^{-1}$, given the WMAP data $d$ (solid black), after marginalizing the
bubble radius.
As explained in the text, the tails of the likelihood are slow to decay, due to a
poorly constrained region of parameter space with small bubble radius.
We illustrate this by showing the likelihood 
calculated using bubbles with a subset of angular sizes:
$\thetabubble < 20^{\circ}$ (blue short-dashed),
and $\thetabubble > 20^{\circ}$ (red long-dashed). \\
{\em Bottom panel:}
Posterior likelihood ${\mathcal L}(\alpha^{\rm ramp}|d)$, obtained from the top panel
by changing variables from $a^{\rm ramp}$ to $\alpha^{\rm ramp} = a^{\rm ramp} (\sin\thetabubble)^{4/3}$.
After this change of variables, the likelihood is narrower and less sensitive to marginalization
over the bubble radius.
Vertical lines are 95\% confidence limits on the amplitude parameter $\alpha^{\rm ramp}$.
The likelihood is consistent with no bubbles ($\alpha^{\rm ramp}=0$).}
  \label{fig:La_ramp}
\end{figure}

Starting from Eq.~(\ref{eqn:bayes_likelihood}) we can marginalize over different parameters
or take slices through the likelihood to calculate the posterior probability distributions for
each parameter.
We begin by setting $a^{\rm step} = 0$, to obtain the 2D likelihood ${\mathcal L}(a^{\rm ramp},r)$
appropriate for the ramp model.
If we now marginalize over the distance parameter $r$, we obtain the 1D likelihood
${\mathcal L}(a^{\rm ramp})$ shown in the top panel of Fig.~\ref{fig:La_ramp}.
Assuming a uniform prior on $a^{\rm ramp}$, the 95\% confidence interval is
$-1.76 \times 10^{-6} < a^{\rm ramp} < 4.27 \times 10^{-6} \;[$Mpc$^{-1}]$.
This constraint is much weaker than would naively be expected from Fig.~\ref{fig:La_ramp}.
The weak constraint arises because bubbles with small angular size are poorly constrained,
even for fairly large values of $a^{\rm ramp}$.
When we marginalize over the angular size to obtain the likelihood ${\mathcal L}(a^{\rm ramp})$,
this leads to tails which are slow to decay.

The broadening of the likelihood caused by the small bubbles is mainly due
to our use of the amplitude parameter $a^{\rm ramp}$, the slope of the initial
curvature perturbation in Mpc$^{-1}$ (as opposed to the peak temperature in the CMB maps, for example).
For a fixed value of $a^{\rm ramp}$, a small bubble corresponds to a much smaller CMB
fluctuation on our sky than a large bubble, and the signal-to-noise is further
diluted by having many small patches on the sky.
This leads to a poorly constrained region in
the two-parameter space $(a^{\rm ramp}, \thetabubble)$ 
where $\thetabubble$ is small and $a^{\rm ramp}$
can be large.

This interpretation of the broadening effect
suggests reparametrizing by replacing the amplitude
parameter $a^{\rm ramp}$ by a variable which is more closely
matched to the statistical significance of the CMB signal.
We define
\be\nonumber
\alpha^{\rm ramp} = a^{\rm ramp} \left(\sin{\thetabubble}\right)^{4/3} \ , \quad \thetabubble = \cos^{-1}{\left(r/D_{dc}\right)}
\ee
where $D_{dc}$ is the comoving distance to last scattering,
and the $4/3$ exponent is empirically chosen so that two bubbles with the same value of
$\alpha$ and different angular sizes have roughly equal statistical significance.
We can now obtain a 1D likelihood ${\mathcal L}(\alpha^{\rm ramp})$ by marginalizing over
the bubble size parameter $\thetabubble$.
This is analogous to our previous marginalization;
note that the Bayesian prior in the new variables
$(\alpha^{\rm ramp},\thetabubble)$ is
$
dP = da\,dr \propto (\sin\thetabubble)^{-1/3} d\alpha \, d\thetabubble \,.
$

The likelihood $\mathcal{L}(\alpha^{\rm ramp})$ is shown
in the bottom panel of Fig.~\ref{fig:La_ramp}.
The $\alpha^{\rm ramp}$ parameter has a narrower distribution than $a^{\rm ramp}$,
and we can achieve a tighter confidence interval.
We obtain the~95\% confidence limits:
\be
-4.66 \times 10^{-8} < a^{\rm ramp}( \sin{\thetabubble})^{4/3} < 4.73 \times 10^{-8} \mbox{ Mpc$^{-1}$}
\ee
which we take to be our ``bottom line'' constraints on the ramp model.

\begin{figure}[t]
  \centering
  \includegraphics[width=71mm, clip=true,trim=0.1cm 0 0 0]{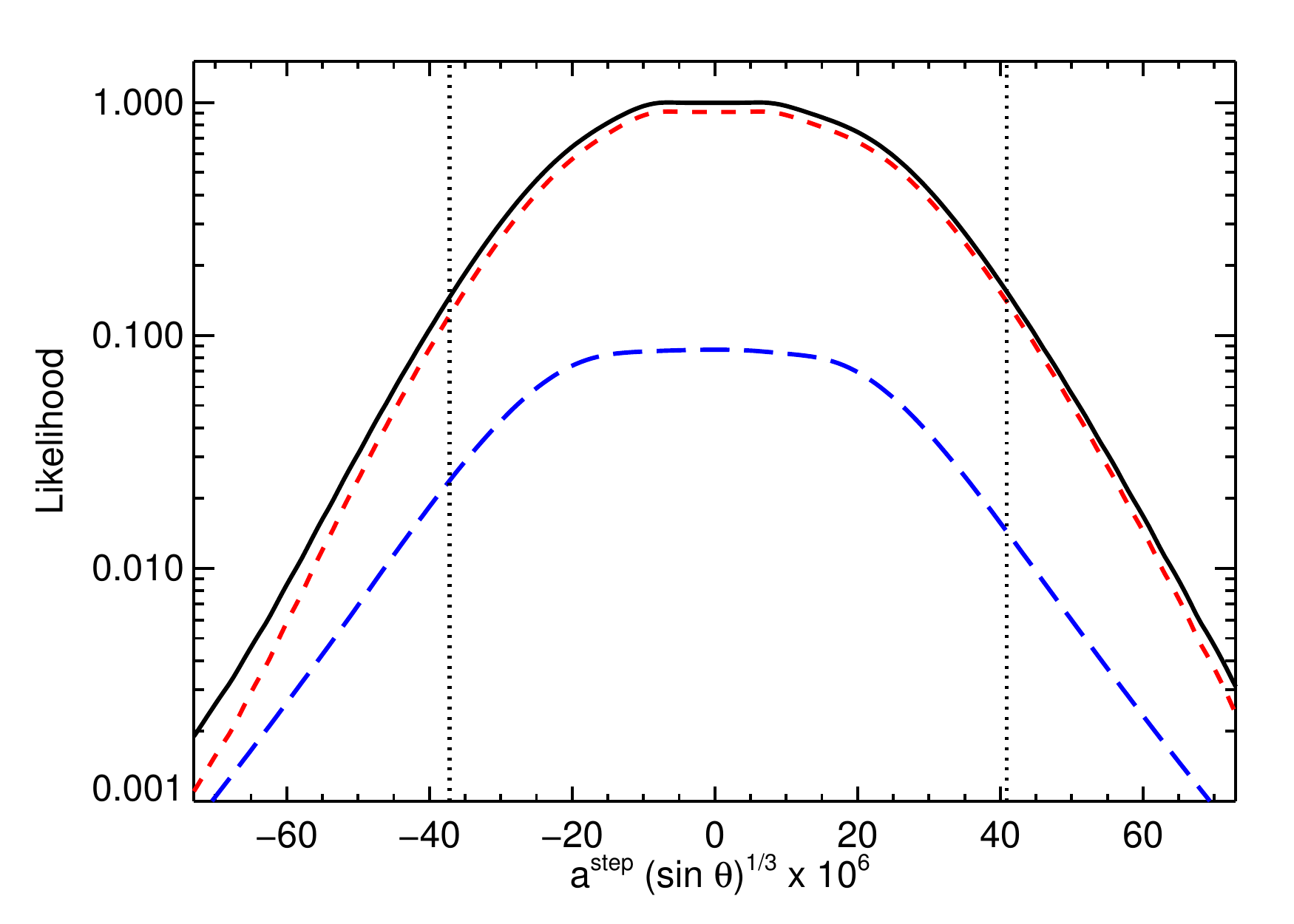}
  \caption{Bayesian analysis of the ``step'' bubble model.
We show the posterior likelihood ${\mathcal L}(\alpha^{\rm step}|d)$ after changing variables
from $a^{\rm step}$ to $\alpha^{\rm step} = a^{\rm step} (\sin\thetabubble)^{1/3}$ to remove degeneracies.
The dashed coloured lines have the same meaning as in Fig.~\ref{fig:La_ramp}.
The likelihood is consistent with no detection of the signal.}
  \label{fig:La_step}
\end{figure}

A similar Bayesian analysis can be performed for the step model.
Considering first the posterior likelihood $\mathcal{L}(a^{\rm step}|d)$ with the bubble radius and
location marginalized, we find that the likelihood has slowly decaying tails leading to a weak
constraint.
Changing variables from $a^{\rm step}$ to $a^{\rm step} (\sin\thetabubble)^{1/3}$,
we find a well-behaved likelihood (Fig.~\ref{fig:La_step}), and the ``bottom line''
95\% confidence limits:
\be
-3.72 \times 10^{-5} < a^{\rm step}( \sin{\thetabubble})^{1/3} < 4.09 \times 10^{-5}
\ee

For both the ramp model and step model, the maximum likelihood amplitude is
very close to zero, much closer than the width of the likelihood.
While this behaviour appears counterintuitive, we explore this phenomenon
in detail in~\cite{technical} and show that it has a natural explanation.
In simulations we find that the maximum likelihood amplitude is nearly zero in
an order-one fraction of the realizations.
The scatter between maximum likelihood estimates of the amplitude, taken over
many simulations, is consistent with Fisher matrix forecasts and roughly equal 
to the width of the likelihood, as expected intuitively.
For the ramp model, simulations with nearly zero maximum
likelihood amplitude often have low quadrupoles, i.e.~the preference for
zero bubble amplitude in the WMAP data is statistically related to the
low quadrupole.
This can be understood intuitively: in realizations with large
quadrupoles, a bubble can cancel large-scale quadrupole power, and so
a nonzero bubble amplitude is preferred by the likelihood.



\begin{figure}[t]
  \centering
  \includegraphics[width=71mm, clip=true,trim=0.1cm 0 0 0]{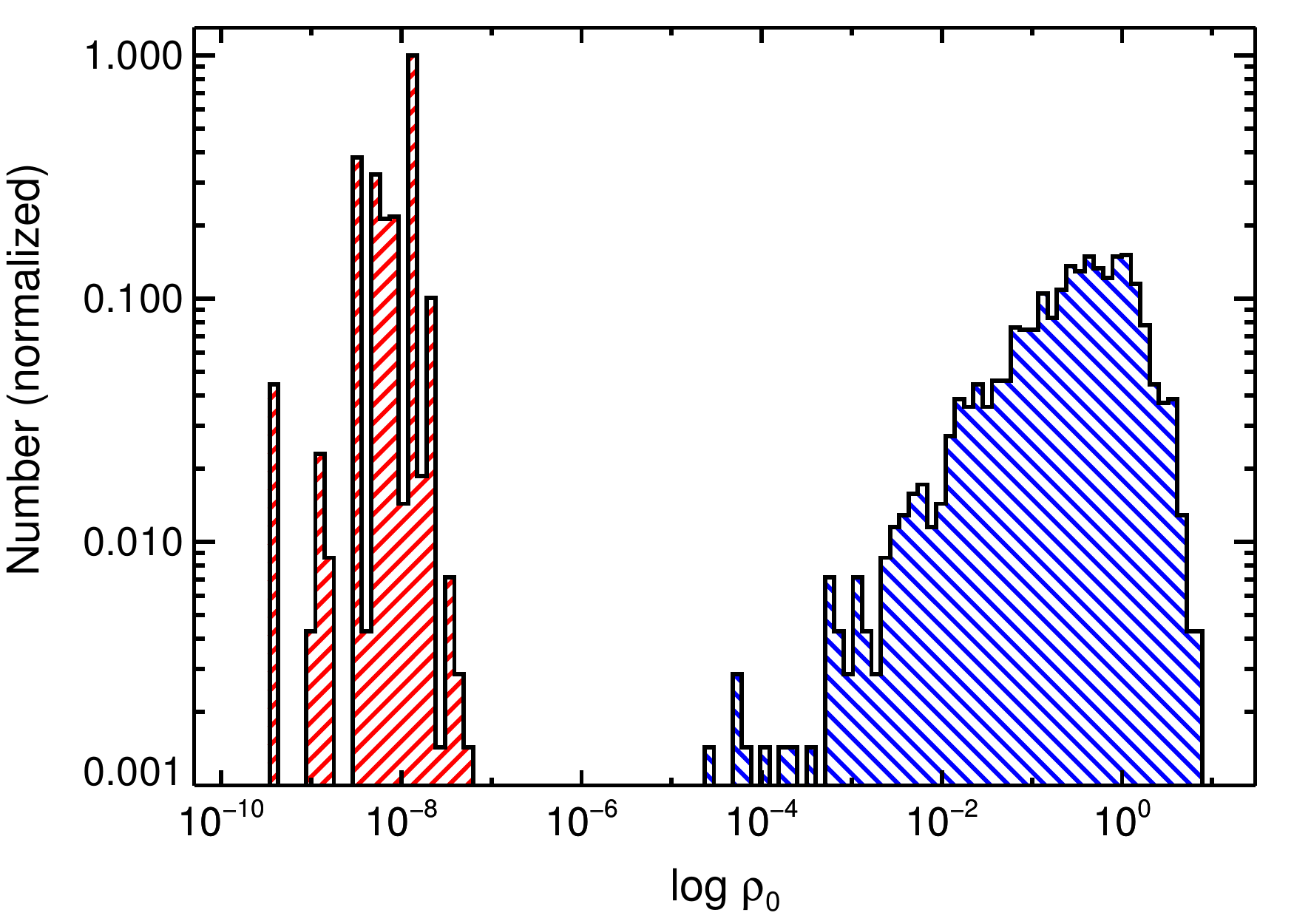}
  \caption[]{Distribution of the likelihood ratio from simulations. An
x-axis value of zero means that the likelihood peaks at $a = 0$.
The red up-shaded distribution is consistent with the likelihood peaking at $a = 0$ within numerical precision.
The WMAP value is consistent with a peak at $a = 0$, which occurs in $51.4\%$ of the
simulations.}
  \label{fig:results_frequentist}
\end{figure}

\smallskip
{\em Frequentist Analysis:} We can also use optimal frequentist statistics
to determine whether a single-bubble model with
$a \ne 0$ gives a significantly better fit to the data than a no-bubble model with $a = 0$.
Frequentist confidence regions are defined by Monte Carlo hypothesis testing and do not
use a prior on bubble amplitude parameters, although we do make use of the theoretical
prior on the bubble radius to improve statistical power.
For testing whether $a=0$ is consistent with the data, the Neyman-Pearson lemma
implies that the optimal frequentist test statistic is the likelihood ratio:
\be
\rho_0(d) = \max_{a} \frac{{\mathcal L}(d|a)}{{\mathcal L}(d|0)}
\ee
We evaluate $\rho_0$ on the WMAP data, and compare it to a histogram of $\rho_0$
values obtained from 5000 Monte Carlo simulations.
The results for the ramp model are shown in Fig.~\ref{fig:results_frequentist};
we find that 48.6\% of the simulations have $\rho_0$ values larger than WMAP, so
WMAP is consistent with $a=0$.

\smallskip
{\em Many bubbles:}
We now consider the possibility that the data contain a large number of low amplitude
bubbles, none of which could be detected individually. This case was first considered
in~\cite{Kozaczuk:2012sx} assuming the Sachs-Wolfe approximation. Theoretically it
is expected that either bubble collisions are unlikely to be present in the data, or that a large number of
collision walls have intersected the last scattering surface~\cite{Kleban:2011pg}.
A large number $N \gg 1$ of independent random bubbles will add a Gaussian signal to the data with power spectrum:
\be
\label{eqn:multi_bub_result}
C^{\rm bub}_\ell = \frac{N \left<a\right>^2}{4 \pi r_{\rm max}} \; \int dr \, b_\ell^2
\ee
where $a$ is the bubble amplitude, which is now a random variable, $N$ is the expected number
of collisions with our Hubble volume, and $b_\ell$ is the harmonic-space profile (defined precisely
in~\cite{technical}) of a bubble at comoving distance $r$.
For the ramp profile, $C_\ell^{\rm bub}$ falls off roughly as $1/\ell^5$, and
a constraint on the amplitude of the spectrum largely comes from a
measurement of the CMB quadrupole.
Since the WMAP quadrupole is measured to be lower than the
best-fit CMB spectrum, we immediately find that there is no evidence for the multi-bubble spectrum
in the WMAP data.
Using the WMAP likelihood code~\cite{Komatsu:2010fb}, we find the upper limit:
\be
\langle N a_{\rm ramp}^2 \rangle^{1/2} \le 6.95 \times 10^{-8} \mbox{ Mpc$^{-1}$ (95\% CL)}
\ee
For the step profile, $C_\ell^{\rm bub}$ falls off roughly as $1/\ell^3$.
The power spectrum constraint comes from a wide range of $\ell$ and we obtain:
\be
\langle N a_{\rm step}^2 \rangle^{1/2} \le 3.52 \times 10^{-4} \mbox{ (95\% CL)}
\ee


\smallskip
{\em Conclusions:} We have searched for evidence that our universe collided with a
bubble universe born out of a nucleation bubble from a phase of false vacuum eternal inflation.
We use an efficient, optimal estimator for detecting the bubble signal in CMB maps
and discuss the technical details in a companion
paper~\cite{technical}. We find no evidence for the bubble signal when applying
our estimator to the WMAP 7-year data, and we place limits on the amplitude of the signal.

The bubble signal comes mainly from low $\ell$ in the ramp model, and from intermediate $\ell$ in the
step model.  Therefore, Planck data is unlikely to improve constraints on the ramp model parameters
(although polarization may help a little~\cite{Chang:2008gj}), but will improve constraints on the step model.
Large-scale structure surveys and 21-cm line surveys can potentially improve limits on both models.


We thank A.~Brown and M.~Kleban. SJO is supported by the US Planck Project, which is funded by the NASA Science Mission
Directorate. LS~is supported by DOE Early Career Award DE-FG02-12ER41854 and the National Science Foundation
under PHY-1068380. KMS is supported at Princeton University by a Lyman Spitzer fellowship, and
at Perimeter Institute by the Government of Canada and the Province of Ontario.
Some of the results have been derived using HEALPix~\cite{Gorski:2004by}. 
We acknowledge the use of the Legacy Archive for Microwave Background Data Analysis (LAMBDA).


\end{document}